\documentclass[conference]{IEEEtran}
\IEEEoverridecommandlockouts
\usepackage{cite}
\usepackage{amsmath,amssymb,amsfonts}
\usepackage{enumitem}
\usepackage{algorithmic}
\usepackage[ruled,vlined]{algorithm2e}
\usepackage{graphicx}
\usepackage{textcomp}
\usepackage{xcolor}
\usepackage{booktabs}
\usepackage{adjustbox}
\PassOptionsToPackage{numbers, sorted}{natbib}
\def\BibTeX{{\rm B\kern-.05em{\sc i\kern-.025em b}\kern-.08em
    T\kern-.1667em\lower.7ex\hbox{E}\kern-.125emX}}

\begin{document}


\title{Joint Partitioning and Placement of \\Foundation Models for Real-Time Edge AI
\thanks{
\noindent
This work was supported by the German Federal Ministry of Research, Technology and Space within the project 6G-life (Grant 16KISK002), by the Bavarian Ministry of Science and the Arts through the project Next Generation AI Computing (gAIn), and by the Bavarian Ministry of Economic Affairs, Regional Development and Energy through the project 6G Future Lab Bavaria.
}
}


\author{%
   \IEEEauthorblockN{%
       Aladin Djuhera\IEEEauthorrefmark{1},
       Fernando Koch\IEEEauthorrefmark{2},
       Alecio Binotto\IEEEauthorrefmark{3} \\
   }
   \IEEEauthorblockA{
       \IEEEauthorrefmark{1}Technical University of Munich, Germany,
       \IEEEauthorrefmark{2}Florida Atlantic University, USA,
       \IEEEauthorrefmark{3}Carl Zeiss AG, Germany\\
       Emails: aladin.djuhera@tum.de, kochf@fau.edu, alecio.binotto@zeiss.com
   }
}

\newcommand{\refig}[1]{Fig.~\ref{#1}}
\newcommand{\refalg}[1]{Alg.~\ref{#1}}

\maketitle


\begin{abstract}
Inference over large-scale foundation models within heterogeneous edge environments necessitates a fundamentally reconfigurable orchestration substrate. 
Static partitioning of model layers presumes temporal stability across compute and network resources, which is  misaligned with the volatility of real-world deployments. 
We introduce a framework in which both the spatial placement and internal segmentation of foundation models are elevated to runtime-resolved constructs. 
The orchestration problem is formalized as a constrained optimization over layer-wise assignments, subject to evolving latency, utilization, and privacy gradients. 
The framework implements reactive inference composition responsive to infrastructural fluctuations by integrating model-aware capacity profiling with dynamic graph re-partitioning and reallocation. 
We introduce architectural and algorithmic components, along with a representative use case in 6G multi-access edge computing.
\end{abstract}



\section{Introduction}
\label{sec:introduction}

Next-generation 6G-enabled networks will need to support a multitude of AI services based on large foundation models (LFM), such as transformer-based large language models (LLM) \cite{llms_telecom}. 
However, deploying LFMs for inference requires significant compute, making their adoption challenging for edge environments \cite{li2024llm}.
\emph{Distributed split inference} (DSI) \cite{karjee2022split} has emerged as a promising approach to alleviate the computational burden. 
It partitions an LFM into multiple segments that are executed sequentially across different nodes, e.g., some executed locally, while heavier segments can be outsourced.
However, such \emph{model splits} are mostly static and predetermined before execution, thus lacking adaptability to dynamic and heterogeneous operational conditions such as fluctuating network reliability or changing node utilization.
Consequently, these approaches lead to suboptimal performance, compromising latency, resource utilization, and quality of service (QoS) guarantees, especially in mission-critical applications \cite{karjee2021split}.
This problem becomes even more acute in \emph{resource-constrained} and \emph{heterogeneous} edge environments, where multiple users rely on accessing shared resources such as multi-access edge compute (MEC), and where data privacy regulations often restrict offloading computations to the cloud.
Such volatility renders any \emph{a priori} static split untenable.

\begin{figure}[t]
    \centering
    \includegraphics[width=0.94\linewidth]{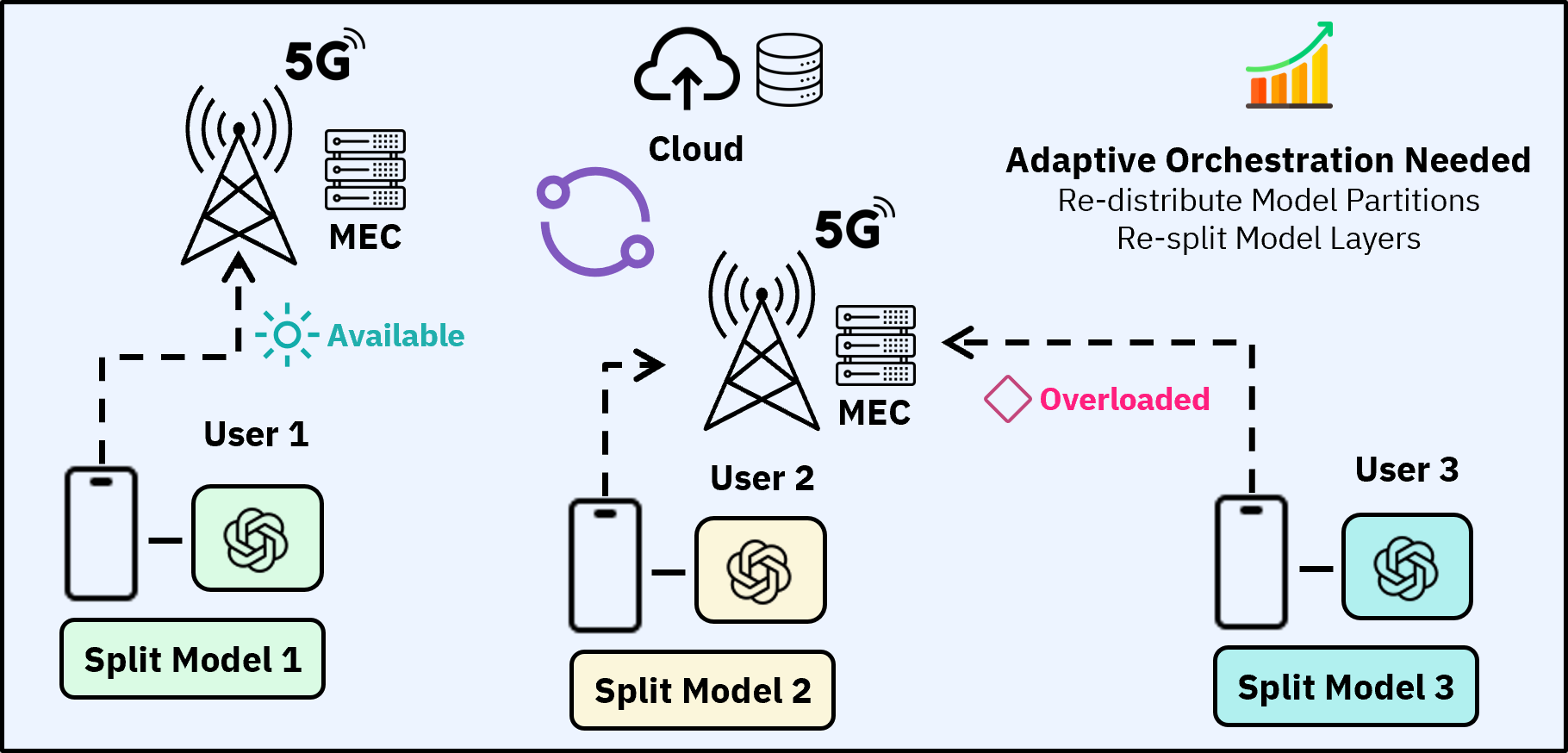}
    \caption{DSI with LFMs requires adaptive orchestration between edge and cloud nodes to guarantee latency, service quality, and efficient node utilization.}
    \label{fig:intro}
    \vspace{-1em}
\end{figure}

In addition, existing inference orchestration frameworks, such as Kubernetes, Ray Serve, InferLine, and KubeEdge, excel at container or micro-batch scheduling but treat LFMs primarily as black boxes, thus lacking mechanisms for runtime layer re-partitioning or privacy-aware placement \cite{vasireddy2023kubernetes}. 
Consequently, the key problem of \emph{joint model-aware partition and placement} under dynamic edge conditions remains open.

In this work, we address this problem by introducing an \emph{adaptive split inference orchestration framework}, extending existing workload orchestrators with domain-specific capabilities specifically tailored for LFMs, such as (multi-modal) LLMs. 
We introduce the following capabilities by leveraging the modular architecture of these models:

\begin{enumerate}
    \item \textbf{Distribution of workloads to edge nodes} with better performance or capacity than the original source node.
    
    \item \textbf{Relocation of split LFM segments} to dynamically optimize resources under changing compute conditions.
    
    \item \textbf{Dynamic LFM re-splitting} to further improve performance and resource utilization when required.
\end{enumerate}

Unlike general-purpose workload schedulers, our framework operates on the \emph{computational graph of the LFM itself}, allowing decisions at the granularity of, e.g., individual transformer blocks. 
This enables QoS-driven re-splitting that commodity orchestrators do not address.
Furthermore, as our solution emphasizes split inference, \emph{privacy} can be implemented as an additional feature at no cost if sensitive LFM layers can be executed locally, which makes reverse engineering data from model weights significantly more challenging for attackers \cite{xu2021privacy}. 
Through this approach, we establish a foundation for \emph{real-time} and \emph{QoS-aware} LFM inference in edge networks, aligning with key 6G objectives of seamless connectivity, low inference latency, and intelligent edge resource management \cite{letaief2021edge}.


\section{Background}
\label{sec:background}

\subsection{Challenges in Edge AI Workload Orchestration}
The current industry norm has been to integrate general-purpose orchestration platforms (e.g., Kubernetes or proprietary MEC orchestrators), which lack mechanisms to dynamically redistribute or reconfigure large model partitions based on real-time changes in \emph{network conditions, node utilization}, or \emph{connectivity} \cite{carrion2022kubernetes}. 
Further, while model-serving stacks such as Triton, InferLine, Ray~Serve, and MLC-Serve introduce batching or replica autoscaling, they still treat neural networks as opaque binaries and cannot \emph{re-partition} a model graph at runtime.
Edge extensions (e.g., KubeEdge and OpenYurt) inherit the same pod-level abstraction, leaving the joint problem of \emph{joint layer splitting and placement}, unaddressed.
As a result:
\begin{itemize}
    \item \textbf{Latency spikes} occur when critical links become congested, delaying real-time applications.
    \item \textbf{Straggler problems} arise when tasks are bottlenecked on overloaded or slower nodes, degrading overall QoS.
    \item \textbf{Resource utilization} becomes imbalanced, either overloading certain nodes or leaving others underutilized, leading to missed service-level agreements (SLAs).
    \item \textbf{Privacy risks} escalate when sensitive data must be offloaded remotely due to inadequate local processing.
\end{itemize}

While current research predominantly focuses on \emph{efficient AI model training} \cite{duan2024efficienttraininglargelanguage, lang2024comprehensive}, the practical challenges of \emph{efficient inference} remain relatively overlooked \cite{zhou2024survey}. Yet, these challenges are increasingly critical for the widespread adoption of LFMs in industrial and commercial scenarios, particularly in future \emph{AI as a Service} (AIaaS)-driven 6G networks \cite{saad2024artificialgeneralintelligenceaginative}.

\subsection{Distributed and Adaptive Split Inference}
DSI partitions a model across different compute locations (e.g., client device, MEC node, cloud), where often lightweight or privacy-sensitive model layers are executed on-device and subsequent layers are offloaded to a remote server \cite{mohammed2020distributed}.

Although DSI enables LFMs to operate closer to data sources, current implementations predominantly employ static splits defined \textit{a priori} based on expected conditions without runtime adaptation. 
This becomes problematic when, for example, a single MEC-enabled base station becomes saturated by various other inference workloads \cite{lin2023pushing} (see \refig{fig:intro}).
While some studies, such as EdgeShard \cite{EdgeShard}, explore collaborative inference setups where a model is shared across edge nodes, these approaches continue to lack dynamic orchestration of model splits and thus cannot effectively respond to real-world changes, such as fluctuating node workloads or intermittent connectivity, resulting in suboptimal latency, inefficient resource utilization, and degraded service quality \cite{hudson2024qos}. 
Thus, in standard implementations, orchestrators cannot dynamically decide to offload additional LFM layers to another edge node or re-split the neural network, leaving significant performance and reliability gains unrealized.
This is especially problematic for heterogeneous compute nodes. 
Thus, a \emph{one-size-fits-all} static partitioning rarely works, as local workloads, performance constraints, and available resources tend to differ \cite{li2024llm}.

However, recent research highlights the benefits of \emph{adaptive split inference} wherein partition points or even partition strategies (e.g., layer reordering) can be reconfigured at runtime to maintain QoS under shifting conditions \cite{chen2024adaptive}. 
This approach, combined with optimal orchestration policies, has the potential to cater to the increasingly demanding AI inference workloads in future AIaaS 6G-enabled networks and edge environments. 

\subsection{Key Design Goals of Adaptive LFM Split Inference}

Practical deployments remain constrained by \emph{static} or \emph{coarse-grained} orchestration mechanisms~\cite{karjee2021split, zhou2019distributing}. 
Thus, current solutions cannot adapt to shifting network or compute conditions, leading to latency spikes, resource imbalances, and potential SLA and QoS violations~\cite{li2019learning}. 
Meanwhile, next-generation 6G network architectures will further exacerbate the complexity of distributing large-scale inference workloads across heterogeneous edge topologies to support various commercial and operational AIaaS applications~\cite{llms_telecom}. 
Hence, an \emph{adaptive split inference} framework will be required that:

\begin{enumerate}
    \item dynamically reconfigures the partition of LFM layers among edge and cloud compute nodes,
    \item exploits real-time profiling of resource availability,
    \item preserves data privacy by keeping computation local, and
    \item ensures consistent, QoS/SLA-compliant performance.
\end{enumerate}

\noindent
In the next section, we propose a novel orchestration method that closes this gap by intelligently managing LFM inference across edge compute infrastructures.

\section{Proposed Adaptive Orchestration Framework}
\label{sec:proposal}

\begin{figure*}[ht]
    \centering
    \includegraphics[width=1\textwidth]{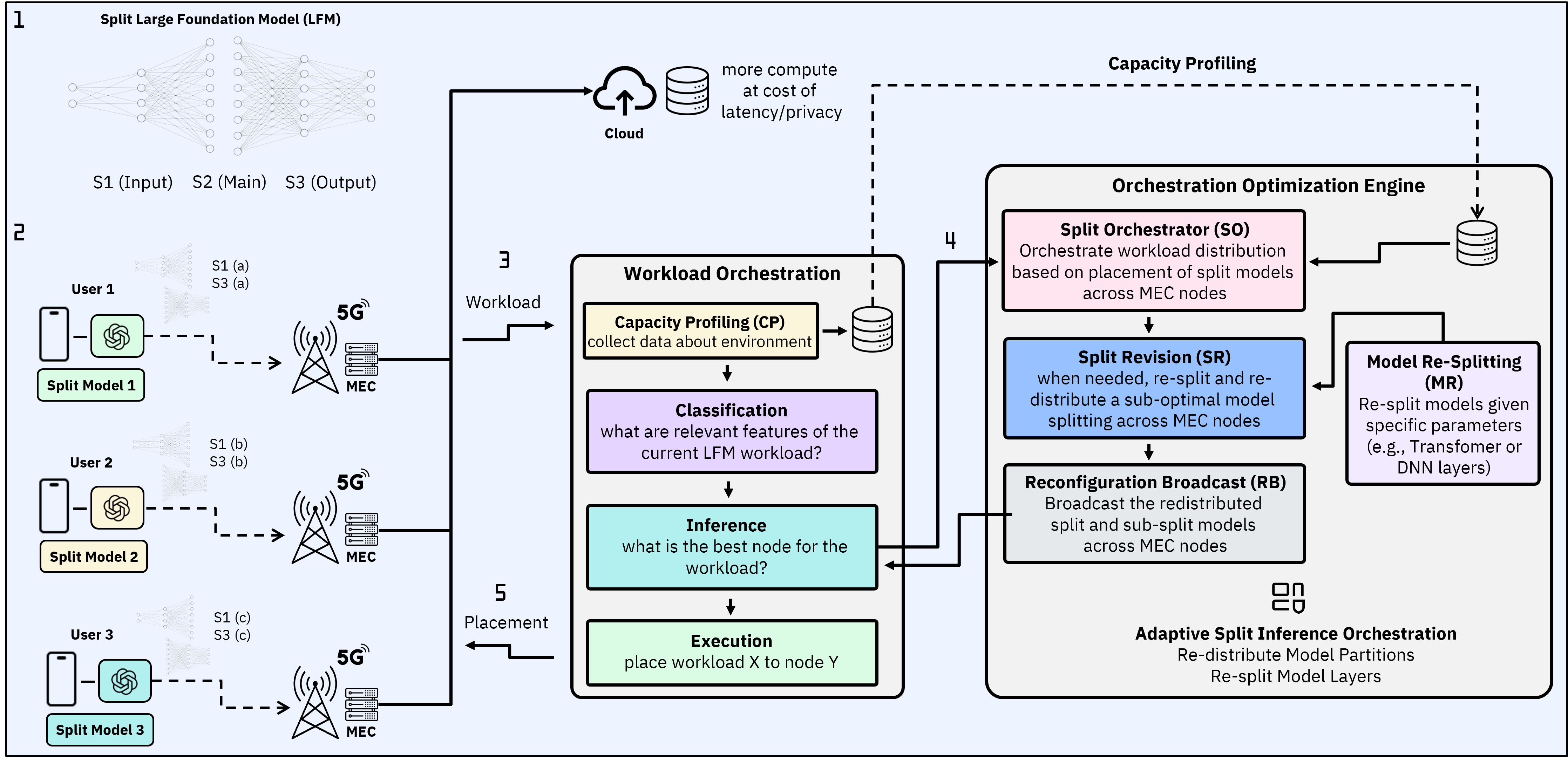}
    \caption{Reference architecture of the proposed adaptive split inference orchestration. Sub-split models (S1, S2, S3) are deployed across edge/cloud nodes, while a central orchestrator, guided by real-time capacity profiling, re-splits and reconfigures workloads on demand to meet QoS and privacy constraints.}
    \label{fig:overview}
\end{figure*}

We propose an \emph{adaptive split inference orchestration} framework that dynamically manages LFM partitions between nodes. 
\refig{fig:overview} depicts a possible realization of this framework in a 5G/6G-MEC deployment, including components for monitoring, decision-making, model partitioning, and reconfiguration. We outline a detailed reference architecture as follows.

\subsection{Reference Architecture}

Our framework orchestrates on-demand allocation and re-allocation of LFM partitions under evolving operational conditions via the following core modules:

\begin{enumerate}
    \item \textbf{Monitoring \& Capacity Profiling (CP):}
    Collects real-time metrics from edge nodes and the network such as CPU/GPU utilization, memory usage, bandwidth, and latency. These metrics guide the orchestrator in partition placement and corresponding re-splitting decisions.

    \item \textbf{Adaptive Orchestrator (AO):}
    Acts as the decision-making engine by evaluating whether to:
    \begin{itemize}
        \item \emph{Keep} the current split (no changes).
        \item \emph{Redistribute} sub-splits across underutilized nodes.
        \item \emph{Fully re-split} the model.
    \end{itemize}
    These decisions are informed by constraints like node capacity, privacy requirements, and expected QoS.
    
    \item \textbf{Split Revision (SR):}
    Implements the logic to re-partition the LFM and may use heuristic, rule-based, or learning-based strategies to identify improved splits.

    \item \textbf{Reconfiguration Broadcast (RB):}
    Broadcasts new model partitions or sub-partitions ($S_{1a}, S_{1b}$, etc.) to the selected nodes and updates local or remote orchestrators.
\end{enumerate}

Our approach dynamically adapts split inference to fluctuating conditions while maintaining strict QoS and privacy requirements by combining these modules. 
The next subsections formalize the system model, define constraints, and describe the orchestration workflow for LFMs in detail.

\subsection{Notation and System Model}

We define key terminologies and orchestration concepts that underlie our adaptive split inference framework as follows. 

\begin{itemize}
    \item \textbf{Compute Nodes.}
    Let $\mathcal{N} = \{1, 2, \ldots, n\}$ denote the set of $n$ edge nodes, and let $c$ refer to a cloud node. 
    Each node $j \in \mathcal{N} \cup \{c\}$ has capacities for inference at time $t$:
    \begin{equation}
        \text{CP}(n_j, t) = \{\mathrm{CPU}_j(t), \mathrm{GPU}_j(t), \mathrm{Mem}_j(t), \mathrm{NetCap}_j(t) \} .
    \end{equation}

    \item \textbf{Partitioning.}
    Consider an LFM \(\mathcal{M}\) segmented into $k$ consecutive model layers:
    \begin{equation}
        S \;=\; \{S_1, S_2, \dots, S_k\}.
    \end{equation}
    
    Typically, $S_1$ handles raw (potentially private) data and $S_k$ generates the final outputs. 
    Intermediate segments $S_2, \dots, S_{k-1}$ often encompass the bulk of computation. 
    A three-split example $\{S_1, S_2, S_3\}$ might place $S_1, S_3$ on a local edge node (for privacy where user data is translated into/from features) and offload the compute-intensive $S_2$ to a more capable node. 
    Depending on the specific LFM architecture, splits may either be configured as self-contained building blocks or individual layers (e.g., from deep or convolutional neural networks) \cite{karjee2021split}.

    \item \textbf{Inference Requests.}
    Inference tasks arrive as requests \(\{r_1, r_2, \dots\}\), each with an associated workload \(\mathcal{W}_r\). 
    At a high level, each request utilizes the same partitions \(\{S_1, \ldots, S_k\}\), but may require separate scheduling decisions depending on QoS constraints or capacity.

    \item \textbf{Decision Variables.}
    For convenience, we define a binary placement matrix \(\mathbf{x} = [x_{i,j}]\), where \(x_{i,j} = 1\) indicates partition \(S_j\) is assigned to node \(n_i\) and \(x_{c,j} = 1\) indicates assignment to the cloud node \(c\). 
    Each column corresponds to a partition and each row to a node in \(\mathcal{N}\cup\{c\}\). 
    When multiple requests are considered, either the same \(\mathbf{x}\) can be reused if the system enforces a single partition, or a time/index extension can be used (e.g., \(\mathbf{x}_{r}\) for each \(r\)).
\end{itemize}

With that, we define the optimization objective as follows.

\ 

\noindent
\textbf{Objective Function.}
We aim to minimize the high-level cost:
\begin{equation}
    \Phi(\mathbf{x}, \mathcal{C}(t))
    \;=\; 
    \alpha\,\mathcal{L}\bigl(\mathbf{x}, \mathcal{C}(t)\bigr)
    +\;
    \beta\,\mathcal{U}\bigl(\mathbf{x}, \mathcal{C}(t)\bigr)
    \nonumber
    +\;
    \gamma\,\mathcal{P}\bigl(\mathbf{x}, \mathcal{C}(t)\bigr),
    \label{eq:cost_function}
\end{equation}
where:
\begin{itemize}
    \item \(\mathcal{L}\) measures inference latency, including data transfer,
    \item \(\mathcal{U}\) captures resource usage imbalance or node overload,
    \item \(\mathcal{P}\) penalizes privacy violations (e.g., placing sensitive partitions on untrusted nodes), and
    \item \(\alpha, \beta, \gamma \ge 0\) weigh the relative importance of latency, resource usage, and privacy, respectively.
\end{itemize}
Here, \(\mathcal{C}(t)\) encapsulates the system state at time \(t\), including node capacities, network bandwidths, and any QoS/SLA requirements. 
In scenarios with concurrent requests, \(\Phi\) can be extended to the sum or average cost across all requests. 
In addition, to ensure valid assignments, we impose constraints:

\begin{enumerate}
    \item \textbf{Unique Assignment.}
    Each partition \(S_j\) must be placed on exactly one node:
    \begin{equation}
        \sum_{i \in \mathcal{N}} x_{i,j} \;+\; x_{c,j} = 1,
        \quad
        \forall j \in \{1,\dots,k\}.
    \end{equation}
    
    \item \textbf{Capacity Limits.}
    For each node \(n_i \in \mathcal{N}\), the sum of resource loads from its assigned partitions cannot exceed that node’s capacity:
    \begin{equation}
        \sum_{j=1}^k \mathrm{load}\bigl(S_j\bigr)\,x_{i,j} 
        \;\;\le\;\; 
        \mathrm{capacity}(n_i, t).
    \end{equation}

    \item \textbf{Privacy Constraints.}
    Partitions handling sensitive data (e.g., \(S_1\)) must remain on trusted nodes:
    \begin{equation}
        x_{i,j} = 0, 
        \quad
        \text{if } n_i \notin \mathrm{trustedSet} \;\wedge\; \bigl(S_j\text{ is privacy-critical}\bigr).
    \end{equation}
    
\end{enumerate}

Further, if LFM layer boundaries can be modified (e.g., subdividing $S_2$ into $\{S_{2a}, S_{2b}\}$, as for example in neural network layers, transformer embeddings, attention layers, etc.), we may treat the set of partitions $S$ itself as part of the optimization. Herewith, we define the split revision as follows. \\

\noindent
\textbf{Split Revision (SR).}
Let \(\Omega\) denote the set of all valid splitting schemes. The orchestrator aims to solve
\begin{equation}
\min_{\;S \in \Omega,\;\mathbf{x}} 
\;\;
\Phi\!\bigl(\mathbf{x}, S, \mathcal{C}(t)\bigr)
\label{eq:splitrevision}
\end{equation}
to find an optimal split \(S^*\) and assignment \(\mathbf{x}^*\) that minimizes the overall cost subject to the constraints above. This allows partitions and assignments to adapt dynamically to shifts in resource availability and workload demands.

\subsection{Orchestration Workflow}

\refalg{alg:workflow} outlines the main orchestration steps. 
The workflow begins by deploying a \emph{baseline} partition (e.g., $\{S_1, S_2, S_3\}$) among a set of nodes. 
The system then continuously monitors resource usage and performance metrics to trigger adjustments.

\begin{enumerate}
    \item \textbf{Initial Deployment.} Perform a static partitioning of the model based on coarse performance estimates (e.g., place $S_1, S_3$ locally for privacy and put $S_2$ on a cloud node~$c$).

    \item \textbf{Continuous Monitoring.} The CP module collects real-time metrics $\mathrm{CP}(n_j, t)$ and calculates an \emph{environment state} $\mathbf{E}(t)$ that captures fluctuations in node utilization.

    \begin{algorithm}[t]
    \small
    \caption{Adaptive Split Orchestration Workflow}
    \label{alg:workflow}
    \DontPrintSemicolon
    \KwIn{
      (i) Initial partitioning $\{S_1,\dots,S_P\}$, 
      (ii) baseline mapping $d_0$, 
      (iii) monitoring intervals $\Delta t$, 
      (iv) trigger‐threshold vector $\Theta=\{L_{\max},U_{\max},B_{\min},T_{\text{cool}}\}$
    }
    \SetAlgoLined
    
    \textbf{Initialize:} Deploy baseline split $(S_1,\ldots,S_P)$ across nodes as per $d_0$.\\
    Set $t_{\mathrm{last}}\gets -\infty$.\\
    
    \For{\emph{each monitoring cycle} $t \leftarrow 0, \Delta t, 2\Delta t, \ldots$}{
      Collect environment metrics $\mathbf{E}(t)$ via Monitoring \& CP.\\
      $\mathit{reconf}\gets \texttt{ShouldReconfigure}(\mathbf{E}(t),\Theta)$.\\
      \uIf{(\emph{trigger condition is met, e.g., high latency, node overload, etc.}) \textbf{and} $\mathit{reconf}$}{
        Evaluate feasible mappings $\{d^\prime\}$ given current partitions.\\
        Optionally call Model Re-Splitting to produce new partitions $\{S_i^*\}$.\\
        Determine best mapping $\hat{d} = \underset{d^\prime}{\arg\min}\,\mathcal{C}(d^\prime)$.\\
        \If{$\hat{d}\neq d_t$ \textbf{and} $t - t_{\mathrm{last}}\ge T_{\text{cool}}$}{
           Broadcast reconfiguration to all affected nodes via RB.\\
           $t_{\mathrm{last}}\gets t$; \quad $d_{t+\Delta t}\gets \hat{d}$.\\
        }
      }
      Resume inference under current assignment $d_{t+\Delta t}$.\\
    }
    \end{algorithm}
    
    \item \textbf{Adaptive Decisions.}
    Based on the updated system states $\mathcal{C}(t), \mathbf{E}(t)$, the adaptive orchestrator continuously evaluates whether to \emph{keep the current split}, \emph{redistribute sub-splits} (reassigning some partitions $S_j$ from node $n_i$ to $n_{i'}$ by adjusting $\mathbf{x}$ without altering the partition boundaries), or \emph{perform full re-splitting} (to obtain a better partition set $S^*$ via the SR module if incremental changes are insufficient or new privacy constraints arise). More formally: 
    \begin{itemize}
        \item The adaptive orchestrator evaluates whether the current partition mapping $d_t$ remains optimal under $\mathbf{E}(t)$. For each request~$r$, the orchestrator checks:
        \begin{equation}
            \mathcal{C}(d_t) \;\overset{?}{\leq}\; \mathcal{C}(d^\prime)
            \quad
            \forall \; \text{feasible } d^\prime.
        \end{equation}
    
        \item If needed, the SR module modifies the set of partitions $\{S_1,\ldots,S_P\}$ (e.g., subdividing a large block $S_2$ into new split configurations $\{S_{2a}, S_{2b}\}$), i.e.
        \begin{equation}
            \hat{d} \;=\; \underset{d \,\in\, \mathcal{D}(\text{new splits})}{\operatorname{argmin}} \,\mathcal{C}(d),
        \end{equation}
        subject to constraints (e.g., compute, network, privacy).
    \end{itemize} 

    \item \textbf{Reconfiguration Broadcast (RB).} Once a decision is made, the \emph{RB} module disseminates the updated assignment $\mathbf{x}^*$ or partition set $S^*$ to relevant nodes, ensuring the new configuration is deployed consistently.

    \item \textbf{Execution.} Inference resumes with the updated partition assignment $\hat{d}$. The orchestrator continues to monitor performance, forming a feedback loop and allowing the system to adapt further as conditions evolve.

\end{enumerate}

\noindent
\textbf{Trigger Conditions and Decision Logic.}
Table~\ref{tab:thr} summarizes the runtime metrics that feed the function \emph{ShouldReconfigure}(E(t),$\Theta$) in \refalg{alg:workflow}.
A reconfiguration is invoked if \emph{any} of the following holds for a monitoring window of length $\Delta t$:

\begin{enumerate}
  \item \textbf{Latency threshold.}  The exponentially weighted moving average (EWMA) of end-to-end inference latency $\bar{L}(t, \Delta t)$ exceeds $L_{\max}$, i.e., $\bar{L}(t,\Delta t) > L_{\max}$.
  
  \item \textbf{Utilisation threshold.}  The maximum node utilization exceeds $U_{\max}$, i.e., $\max_{n \in \mathcal{N}} U_n(t) > U_{\max}$.
  
  \item \textbf{Bandwidth drop.}  The minimum available bandwidth drops below $B_{\min}$, i.e., $\min_{(i,j) \in \mathcal{L}} B_{ij}(t) < B_{\min}$.
\end{enumerate}

\vspace{1ex}
\begin{table}[t]
\centering
\caption{Monitored metrics and default trigger thresholds.}
\label{tab:thr}
\begin{tabular}{lcc}
\toprule
\textbf{Metric} & \textbf{Symbol} & \textbf{Empirical Value} \\
\midrule
EWMA latency                             & $L_{\max}$   & 150 ms \\
GPU/CPU utilization                      & $U_{\max}$   & 0.85    \\
Available link bandwidth (edge-to-edge) & $B_{\min}$   & 50 Mbps \\
Time-to-reconfigure cool-down            & $T_{\text{cool}}$ & 30 s \\
\bottomrule
\end{tabular}
\end{table}

\noindent
If multiple triggers fire simultaneously, the system first attempts \emph{placement migration}. 
If that cannot meet all constraints, the \emph{split-revision} module is invoked. 
Reconfigurations are further rate-limited by $T_{\text{cool}}$ to prevent thrashing.

\ 

In practice, such an orchestration loop can be integrated into existing platforms (e.g., extending Kubernetes with a custom controller for re-splitting). 
Partitioning decisions may then rely on traditional heuristics (e.g., rule-based or greedy approaches) or adopt learning-based schemes (e.g., reinforcement learning) to continuously refine splitting strategies \cite{optimal_ai_splitting, lien2024optimum}.
We leave the investigation of optimal strategies to future work.

\subsection{Privacy and Security Considerations}

A core feature in split inference is the preservation of data privacy by ensuring critical or sensitive operations remain on a trusted device or node. 
Thus, our framework permits a \emph{selective local execution}. 
For example, some LFM blocks, especially those close to the input layer, may handle raw personal or private data. 
By design, these partitions can be configured to remain on the client device or a trusted edge node. 
Formally, if $S_i$ handles private data, we require that
\begin{equation}
        d_t(i) \;\in\; \mathcal{N}_{\text{trusted}} \quad \forall t,
\end{equation}
where $\mathcal{N}_{\text{trusted}} \subseteq \mathcal{N}\cup\{c\}$ is the set of \emph{trusted nodes}. 
Corresponding LFM splits can be obtained according to model architecture, compute resources, and privacy requirements (e.g., measured as layer depth) \cite{optimal_ai_splitting}. 
By leveraging partial LFM layer splits, our orchestration framework thus inherently supports privacy-preserving inference at the edge, ensuring that sensitive data never leaves a trusted domain. 
Furthermore, our framework can be extended with secure encryption and transmission protocols to safeguard intermediate activations (e.g., outputs from $S_1$ that serve as inputs to $S_2$) against malicious entities such as jammers and eavesdroppers \cite{djuhera2024rsfllmjammingresilientframework}.
Beyond inference, our framework readily supports post-hoc safety alignment methods such as SafeMERGE \cite{djuhera2025safemergepreservingsafetyalignment}, which exchange or merge individual model layers with safe counterparts to make them resilient against fine-tuning attacks. By facilitating these layer adaptations, our architecture becomes highly adaptable to a broad spectrum of defense mechanisms.


\section{Expected Performance Results}
\label{sec:expected-results}

We evaluate our adaptive orchestration framework for a simulated 5G-MEC urban scenario based on figures reported
across four public studies \cite{zhang2025amp4ecadaptivemodelpartitioning, tuli2022splitplaceaiaugmentedsplitting, EdgeShard, mudvari2024adaptivecompressionawaresplitlearning}, where we derive \emph{expected KPIs} by combining (i) the ETSI MEC latency model for processing, queuing, and fronthaul/backhaul, as well as (ii) empirical improvements reported by recent split inference works \cite{EdgeShard, chen2024adaptive}.
We choose trigger thresholds as in Table~\ref{tab:thr}.
While we acknowledge that a comparison with baselines like [15] and [24] is valuable, it remains challenging as they lack runtime graph re-splitting capabilities. 
Thus, we leave comprehensive cross-framework benchmarking to future work.

\paragraph{Scenario \& Methodology}
Three MEC nodes, each with an NVIDIA A100 40GB GPU and cloud compute capabilities, serve intermittent LFM inference requests for Llama3 models of 8B parameter size \cite{grattafiori2024llama}. 
The baseline uses a static $\{S_1,S_2,S_3\}$ split, whereas our adaptive orchestrator may migrate segments or re-split \(S_2\) on QoS triggers. 
We sweep backhaul bandwidths \(\{20,50,100,200\}\) Mb/s and compute:
\begin{equation}
    \text{latency}_{\text{static/adaptive}} = \underbrace{T_{\text{proc}} + T_{\text{queue}}}_{\text{ETSI MEC model}} 
+ T_{\text{tx}}(\text{bandwidth}),
\end{equation}

\paragraph{Key Metrics}

\begin{table}[t]
  \centering
  \caption{Expected steady-state KPIs for a 10s monitoring window.}
  \label{tab:expected-kpis}
  \resizebox{0.95\columnwidth}{!}{%
      \begin{tabular}{@{}rcccccc@{}}
        \toprule
        Backhaul & Static-Split & Adaptive & \(\Delta\) & Throughput & GPU \\ 
        (Mb/s)   & Latency (ms) & Latency (ms) & Latency & (\(\times\) Baseline) & Util. \\
        \midrule
         20      & 500          & 200        & –60\%   & 2.1×       & 92\%   \\
         50      & 320          & 150        & –53\%   & 2.0×       & 90\%   \\
        100      & 230          & 120        & –48\%   & 1.9×       & 88\%   \\
        200      & 180          & 110        & –39\%   & 1.8×       & 86\%   \\
        \bottomrule
      \end{tabular}
  }
\end{table}

\begin{figure}[t]
  \centering
  \includegraphics[width=\linewidth]{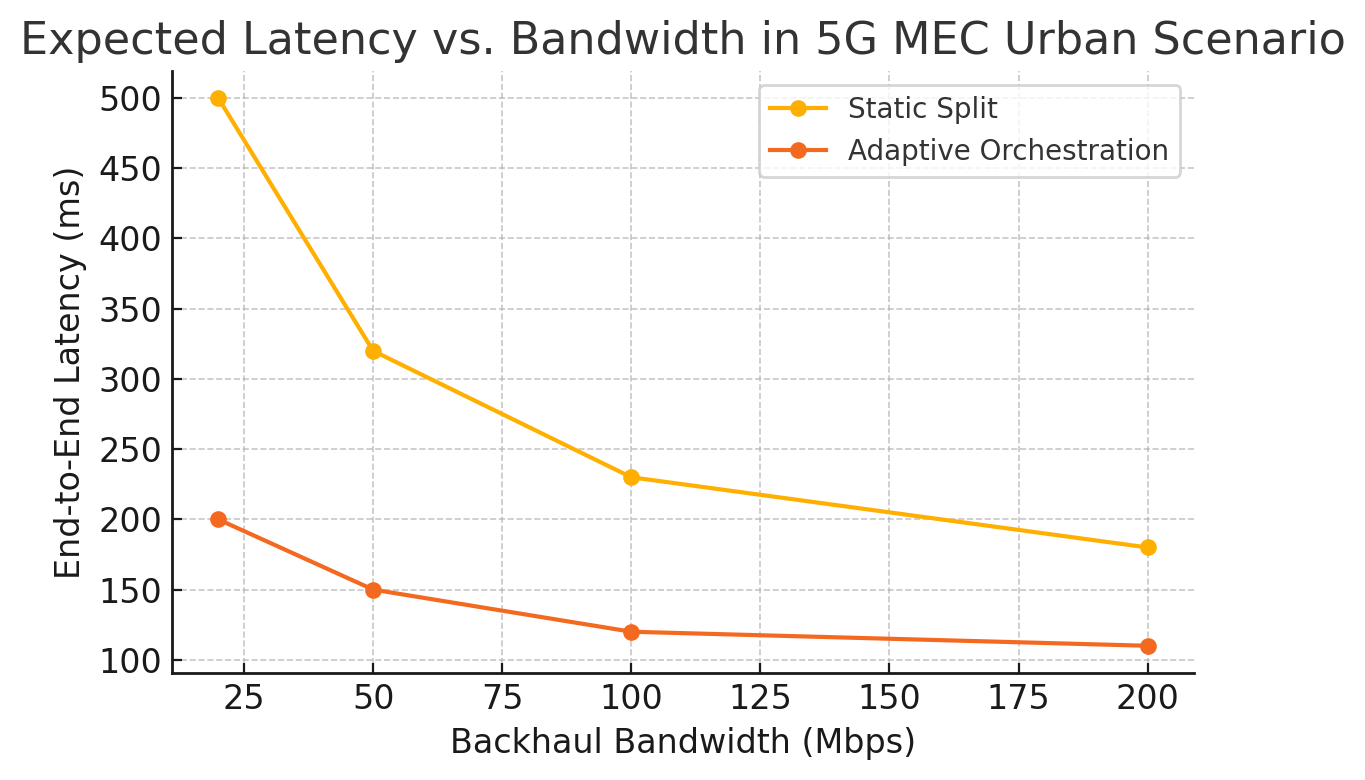}
  \caption{Expected latency vs.\ bandwidth availability in our 5G MEC urban scenario. Adaptive split inference orchestration results in lower total latency.}
  \label{fig:latency}
\end{figure}

In our analysis, we focus on latency as the key performance indicator (KPI). 
Table \ref{tab:expected-kpis} summarizes expected steady-state KPIs averaged over a 10s monitoring window, where deltas inherit margins from related works \cite{zhang2025amp4ecadaptivemodelpartitioning, tuli2022splitplaceaiaugmentedsplitting, mudvari2024adaptivecompressionawaresplitlearning} and scale with the analytical latency model.
\refig{fig:latency} plots the corresponding end-to-end latency versus backhaul bandwidth.
In general, we observe the following improvements:

\begin{itemize}
  \item \textbf{Bandwidth.} Static-split latency falls sharply with bandwidth, whereas adaptive orchestration flattens the curve by migrating or re-splitting to avoid link chokepoints.
  \item \textbf{Diminishing returns.} Above $\sim$100 Mb/s, transport ceases to dominate and gains come from GPU load balancing, narrowing the static-adaptive gap.
  \item \textbf{QoS guarantees.} The 150 ms URLLC bound is met across all loads \emph{only under the adaptive scheme}, highlighting its robustness for V2X and XR services.
  \item \textbf{Resource utilization.} Adaptive re-slicing keeps GPU usage near 90\%, avoiding stragglers and over-provisioning.
\end{itemize}

Adaptive split inference thus consistently outperforms any static configuration. 
The small overhead of monitoring ($\le$10 ms per cycle) and graph rewiring are amortized by hundreds of ms saved per request, yielding a net performance gain an order of magnitude larger than the orchestration cost. 
Consequently, adaptive split inference emerges as the decisive design choice in heterogeneous, bandwidth-variable edge networks.


\section{Conclusions}
\label{sec:conclusions}

This study has introduced an adaptive split inference orchestration framework designed to dynamically manage LFM partitions across heterogeneous edge nodes, establishing a foundation for real-time, QoS-aware, and privacy-preserving AI inference in edge computing environments.
A key innovation of our approach is treating both model partitioning and node placement as dynamic, runtime-resolved variables, rather than fixed pre-conditions.
This design directly addresses the variability in layer-splitting complexity across different LFMs. 
By operating on the underlying computational graph, our framework remains generalizable across diverse architectures without requiring model-specific redesigns.
This unified optimization ensures the system can support latency-sensitive inference and data locality under the volatility of 6G edge environments, offering a scalable substrate that remains extensible to future inference graphs and scheduling objectives.
Further, our framework can be seamlessly integrated with existing orchestration platforms at low cost due to its modular architecture, while maintaining extensibility for future AI-driven optimizations.
Our proposed orchestration model thus optimizes performance, enhances resource efficiency, and fortifies privacy preservation, aligning with emerging objectives in the development of future AI-native 6G and beyond networks.


%
%


\bibliographystyle{IEEEtran} 
\bibliography{mybib}

@misc{djuhera2025safemergepreservingsafetyalignment,
      title={{SafeMERGE: Preserving Safety Alignment in Fine-Tuned Large Language Models via Selective Layer-Wise Model Merging}}, 
      author={Aladin Djuhera and Swanand Ravindra Kadhe and Farhan Ahmed and Syed Zawad and Holger Boche},
      year={2025},
      eprint={2503.17239},
      archivePrefix={arXiv},
      primaryClass={cs.CL},
      url={https://arxiv.org/abs/2503.17239}, 
}

@misc{mudvari2024adaptivecompressionawaresplitlearning,
      title={{Adaptive Compression-Aware Split Learning and Inference for Enhanced Network Efficiency}}, 
      author={Akrit Mudvari and Antero Vainio and Iason Ofeidis and Sasu Tarkoma and Leandros Tassiulas},
      year={2024},
      eprint={2311.05739},
      archivePrefix={arXiv},
      primaryClass={cs.NI},
      url={https://arxiv.org/abs/2311.05739}, 
}

@misc{chen2024adaptive,
      title={{Adaptive Layer Splitting for Wireless LLM Inference in Edge Computing: A Model-Based Reinforcement Learning Approach}}, 
      author={Yuxuan Chen and Rongpeng Li and Xiaoxue Yu and Zhifeng Zhao and Honggang Zhang},
      year={2024},
      eprint={2406.02616},
      archivePrefix={arXiv},
      primaryClass={cs.LG},
      url={https://arxiv.org/abs/2406.02616}, 
}

@misc{tuli2022splitplaceaiaugmentedsplitting,
      title={{SplitPlace: AI Augmented Splitting and Placement of Large-Scale Neural Networks in Mobile Edge Environments}}, 
      author={Shreshth Tuli and Giuliano Casale and Nicholas R. Jennings},
      year={2022},
      eprint={2205.10635},
      archivePrefix={arXiv},
      primaryClass={cs.DC},
      url={https://arxiv.org/abs/2205.10635}, 
}

@misc{zhang2025amp4ecadaptivemodelpartitioning,
      title={{AMP4EC: Adaptive Model Partitioning Framework for Efficient Deep Learning Inference in Edge Computing Environments}}, 
      author={Guilin Zhang and Wulan Guo and Ziqi Tan and Hailong Jiang},
      year={2025},
      eprint={2504.00407},
      archivePrefix={arXiv},
      primaryClass={cs.DC},
      url={https://arxiv.org/abs/2504.00407}, 
}

@article{letaief2021edge,
  title={{Edge Artificial Intelligence for 6G: Vision, Enabling Technologies, and Applications}},
  author={Letaief, Khaled B and Shi, Yuanming and Lu, Jianmin and Lu, Jianhua},
  journal={IEEE journal on selected areas in communications},
  volume={40},
  number={1},
  pages={5--36},
  year={2021},
  publisher={IEEE}
}

@article{xu2021privacy,
  title={{Privacy-Preserving Machine Learning: Methods, Challenges and Directions}},
  author={Xu, Runhua and Baracaldo, Nathalie and Joshi, James},
  journal={arXiv preprint arXiv:2108.04417},
  year={2021}
}

@article{vasireddy2023kubernetes,
  title={{Kubernetes and Docker Load Balancing: State-of-the-Art Techniques and Challenges}},
  author={Vasireddy, Indrani and Ramya, G and Kandi, Prathima},
  journal={International Journal of Innovative Research in Engineering and Management},
  volume={10},
  number={6},
  pages={49--54},
  year={2023}
}

@article{grattafiori2024llama,
  title={{The Llama 3 Herd of Models}},
  author={Grattafiori, Aaron and Dubey, Abhimanyu and Jauhri, Abhinav and Pandey, Abhinav and Kadian, Abhishek and Al-Dahle, Ahmad and Letman, Aiesha and Mathur, Akhil and Schelten, Alan and Vaughan, Alex and others},
  journal={arXiv preprint arXiv:2407.21783},
  year={2024}
}

@article{lien2024optimum,
  title={{Optimum Splitting Computing for DNN Training Through Next Generation Smart Networks: A Multi-Tier Deep Reinforcement Learning Approach}},
  author={Lien, Shao-Yu and Yeh, Cheng-Hao and Deng, Der-Jiunn},
  journal={Wireless Networks},
  volume={30},
  number={3},
  pages={1737--1751},
  year={2024},
  publisher={Springer}
}

@ARTICLE{optimal_ai_splitting,
  author={Li, Xian and Bi, Suzhi},
  journal={IEEE Transactions on Wireless Communications}, 
  title={{Optimal AI Model Splitting and Resource Allocation for Device-Edge Co-Inference in Multi-User Wireless Sensing Systems}}, 
  year={2024},
  volume={23},
  number={9},
  pages={11094-11108},
  doi={10.1109/TWC.2024.3378418}}

@misc{djuhera2024rsfllmjammingresilientframework,
      title={{R-SFLLM: Jamming Resilient Framework for Split Federated Learning with Large Language Models}}, 
      author={Aladin Djuhera and Vlad C. Andrei and Xinyang Li and Ullrich J. Mönich and Holger Boche and Walid Saad},
      year={2024},
      eprint={2407.11654},
      archivePrefix={arXiv},
      primaryClass={cs.LG},
      url={https://arxiv.org/abs/2407.11654}, 
}

@article{hudson2024qos,
  title={{QoS-Aware Edge AI Placement and Scheduling with Multiple Implementations in FaaS-Based Edge Computing}},
  author={Hudson, Nathaniel and Khamfroush, Hana and Baughman, Matt and Lucani, Daniel E and Chard, Kyle and Foster, Ian},
  journal={Future Generation Computer Systems},
  volume={157},
  pages={250--263},
  year={2024},
  publisher={Elsevier}
}

@article{li2024llm,
  title={{LLM Inference Serving: Survey of Recent Advances and Opportunities}},
  author={Li, Baolin and Jiang, Yankai and Gadepally, Vijay and Tiwari, Devesh},
  journal={arXiv preprint arXiv:2407.12391},
  year={2024}
}

@ARTICLE{llms_telecom,
  author={Zhou, Hao and Hu, Chengming and Yuan, Ye and Cui, Yufei and Jin, Yili and Chen, Can and Wu, Haolun and Yuan, Dun and Jiang, Li and Wu, Di and Liu, Xue and Zhang, Charlie and Wang, Xianbin and Liu, Jiangchuan},
  journal={IEEE Communications Surveys \& Tutorials}, 
  title={{Large Language Model (LLM) for Telecommunications: A Comprehensive Survey on Principles, Key Techniques, and Opportunities}}, 
  year={2024},
  volume={},
  number={},
  pages={1-1},
  doi={10.1109/COMST.2024.3465447}}

@ARTICLE{EdgeShard,
  author={Zhang, Mingjin and Shen, Xiaoming and Cao, Jiannong and Cui, Zeyang and Jiang, Shan},
  journal={IEEE Internet of Things Journal}, 
  title={{EdgeShard: Efficient LLM Inference via Collaborative Edge Computing}}, 
  year={2024},
  volume={},
  number={},
  pages={1-1},
  doi={10.1109/JIOT.2024.3524255}}

@article{carrion2022kubernetes,
  title={{Kubernetes Scheduling: Taxonomy, Ongoing Issues and Challenges}},
  author={Carri{\'o}n, Carmen},
  journal={ACM Computing Surveys},
  volume={55},
  number={7},
  pages={1--37},
  year={2022},
  publisher={ACM New York, NY}
}

@article{lin2023pushing,
  title={{Pushing Large Language Models to the 6G Edge: Vision, Challenges, and Opportunities}},
  author={Lin, Zheng and Qu, Guanqiao and Chen, Qiyuan and Chen, Xianhao and Chen, Zhe and Huang, Kaibin},
  journal={arXiv preprint arXiv:2309.16739},
  year={2023}
}

@misc{saad2024artificialgeneralintelligenceaginative,
      title={{Artificial General Intelligence (AGI)-Native Wireless Systems: A Journey Beyond 6G}}, 
      author={Walid Saad and Omar Hashash and Christo Kurisummoottil Thomas and Christina Chaccour and Merouane Debbah and Narayan Mandayam and Zhu Han},
      year={2024},
      eprint={2405.02336},
      archivePrefix={arXiv},
      primaryClass={cs.AI},
      url={https://arxiv.org/abs/2405.02336}, 
}

@article{zhou2024survey,
  title={{A Survey on Efficient Inference for Large Language Models}},
  author={Zhou, Zixuan and Ning, Xuefei and Hong, Ke and Fu, Tianyu and Xu, Jiaming and Li, Shiyao and Lou, Yuming and Wang, Luning and Yuan, Zhihang and Li, Xiuhong and others},
  journal={arXiv preprint arXiv:2404.14294},
  year={2024}
}

@inproceedings{lang2024comprehensive,
  title={{A Comprehensive Study on Quantization Techniques for Large Language Models}},
  author={Lang, Jiedong and Guo, Zhehao and Huang, Shuyu},
  booktitle={IEEE ICAIRC},
  year={2024},
}

@misc{duan2024efficienttraininglargelanguage,
      title={{Efficient Training of Large Language Models on Distributed Infrastructures: A Survey}}, 
      author={Jiangfei Duan and Shuo Zhang and Zerui Wang and others},
      year={2024},
      eprint={2407.20018},
      archivePrefix={arXiv},
      primaryClass={cs.DC},
      url={https://arxiv.org/abs/2407.20018}, 
}

@INPROCEEDINGS{mohammed2020distributed,
  author={Mohammed, Thaha and Joe-Wong, Carlee and Babbar, Rohit and Francesco, Mario Di},
  booktitle={IEEE INFOCOM 2020 - IEEE Conference on Computer Communications}, 
  title={{Distributed Inference Acceleration with Adaptive DNN Partitioning and Offloading}}, 
  year={2020},
  volume={},
  number={},
  pages={854-863},
  doi={10.1109/INFOCOM41043.2020.9155237}}

@article{li2019learning,
  title={{Learning-Aided Computation Offloading for Trusted Collaborative Mobile Edge Computing}},
  author={Li, Yuqing and Wang, Xiong and Gan, Xiaoying and Jin, Haiming and Fu, Luoyi and Wang, Xinbing},
  journal={IEEE Transactions on Mobile Computing},
  volume={19},
  number={12},
  pages={2833--2849},
  year={2019},
  publisher={IEEE}
}

@article{karjee2022split,
author = {Jyotirmoy Karjee and Praveen {Naik S} and Kartik Anand and Vanamala N. Bhargav},
title = {{Split computing: DNN Inference Partition with Load Balancing in IoT-Edge Platform for Beyond 5G}},
journal = {Measurement: Sensors},
volume = {23},
pages = {100409},
year = {2022},
issn = {2665-9174},
doi = {https://doi.org/10.1016/j.measen.2022.100409}
}

@inproceedings{karjee2021split,
  title={{Split Computing: Dynamic Partitioning and Reliable Communications in IoT-Edge for 6G Vision}},
  author={Karjee, Jyotirmoy and Anand, Kartik and Bhargav, Vanamala Narasimha and Naik, Praveen S and Dabbiru, Ramesh Babu Venkat and Srinidhi, N},
  booktitle={2021 8th International Conference on Future Internet of Things and Cloud (FiCloud)},
  year={2021},
}

@inproceedings{zhou2019distributing,
  title={{Distributing Deep Neural Networks with Containerized Partitions at the Edge}},
  author={Zhou, Li and Wen, Hao and Teodorescu, Radu and Du, David HC},
  booktitle={2nd USENIX Workshop on Hot Topics in Edge Computing},
  year={2019}
}


\end{document}